\documentclass[conference]{IEEEtran}
\IEEEoverridecommandlockouts

\usepackage{amsmath,amssymb,amsfonts}
\usepackage[T1]{fontenc}
\usepackage{graphicx}
\usepackage{textcomp}
\usepackage{booktabs}
\usepackage{multirow}
\usepackage{enumitem}
\usepackage{subcaption}
\usepackage{longtable}  
\usepackage{xcolor}
\usepackage{tikz}

\usepackage{svg}

\usepackage{tikz}

\newcommand{\kh}[1]{\textcolor{blue}{#1}}
\renewcommand{\kh}[1]{#1}

\usepackage[
 backend=biber
,style=ieee
,minbibnames=1
,maxbibnames=2
,mincitenames=1
,maxcitenames=2
,eprint=true
,doi=true
,isbn=false
,url=true
]{biblatex}
\bibliography{ref} 
\AtBeginBibliography{\footnotesize}

\DeclareSourcemap{
  \maps{
    \map{
      \step[fieldset=month, null]
      \step[fieldset=address, null]
      \step[fieldset=location, null]
      \step[fieldset=publisher, null]
      \step[fieldset=url, null]
      \step[fieldset=isbn, null]
      \step[fieldset=editor, null]
    }
  }
}

\usepackage{algorithm}
\usepackage[noend]{algpseudocode}
\algnewcommand{\LineComment}[1]{\State \(\triangleright\) #1}
\algnewcommand{\algorithmicand}{\textbf{ and }}
\algnewcommand{\algorithmicor}{\textbf{ or }}
\algnewcommand{\OR}{\algorithmicor}
\algnewcommand{\AND}{\algorithmicand}
\algnewcommand{\var}{\texttt}

\usepackage[
hidelinks,
  colorlinks=true,
  citecolor=blue,
  pdfstartview=Fit,
  breaklinks=true
]{hyperref}

\usepackage{cleveref}
\Crefname{figure}{Fig.}{Figs.}

\usepackage{flushend}

\begin{document}

\title{Taming Volatility: Stable and Private QUIC Classification with Federated Learning}

% Robust Federated Learning for QUIC Traffic Classification in Dynamic Network Environments

% Privacy-Preserving QUIC Classification at Scale: A Federated Learning Approach

% Privacy-Preserving QUIC Traffic Classification Using Federated Learning

\author{
    \IEEEauthorblockN{Richard Jozsa\IEEEauthorrefmark{1}, Karel Hynek\IEEEauthorrefmark{2}\IEEEauthorrefmark{5}, and Adrian Pekar\IEEEauthorrefmark{1}\IEEEauthorrefmark{3}\IEEEauthorrefmark{4}}
    \IEEEauthorblockA{
        \IEEEauthorrefmark{1}Budapest University of Technology and Economics, M\H{u}egyetem rkp. 3., H-1111 Budapest, Hungary\\
        \IEEEauthorrefmark{2}Czech Technical University in Prague, 
        Faculty of Information Technology, 
        Czech Republic\\
        \IEEEauthorrefmark{3}HUN-REN-BME Information Systems Research Group, Magyar Tud\'{o}sok krt. 2, 1117 Budapest, Hungary\\
        \IEEEauthorrefmark{4}CUJO LLC, Budapest, Hungary\\
        \IEEEauthorrefmark{5}CESNET, a.l.e., Prague, Czech Republic\\
        Email: richard.jozsa@edu.bme.hu, hynekkar@fit.cvut.cz, apekar@hit.bme.hu
    }
}

% Department of Networked Systems and Services, Faculty of Electrical Engineering and Informatics,\\ 

\maketitle

\begin{abstract}
Federated Learning (FL) is a promising approach for privacy-preserving network traffic analysis, but its practical deployment is challenged by the non-IID nature of real-world data. While prior work has addressed statistical heterogeneity, the impact of temporal traffic volatility—the natural daily ebb and flow of network activity—on model stability remains largely unexplored. This volatility can lead to inconsistent data availability at clients, destabilizing the entire training process.
In this paper, we systematically address the problem of temporal volatility in federated QUIC classification. We first demonstrate the instability of standard FL in this dynamic setting. We then propose and evaluate a client-side data buffer as a practical mechanism to ensure stable and consistent local training, decoupling it from real-time traffic fluctuations. Using the real-world CESNET-QUIC22 dataset partitioned into 14 autonomous clients, we then demonstrate that this approach enables robust convergence.
% 
% In this paper, we systematically address the problem of temporal volatility in federated QUIC classification. We propose the application of a client-side data buffer as a practical mechanism to ensure stable and consistent local training, decoupling it from real-time traffic fluctuations. Using the real-world CESNET-QUIC22 dataset partitioned into 14 autonomous clients, we demonstrate that this approach enables robust convergence.
% 
Our results show that a stable federated system achieves a 95.2\% F1 score, a mere 2.3 percentage points below a non-private centralized model. This work establishes a blueprint for building operationally stable FL systems for network management, proving that the challenges of dynamic network environments can be overcome with targeted architectural choices.
\end{abstract}


\begin{tikzpicture}[remember picture,overlay]
\node[anchor=north, align=center, text=black, font=\small\itshape, yshift=-.6cm] at (current page.north) {This version of the paper was accepted for presentation at the 2025 International Conference on Network and Service Management (CNSM 2025)};
\end{tikzpicture}

\section{Introduction}

Network traffic classification \kh{(TC)} has become increasingly challenging with the widespread adoption of encrypted protocols, particularly QUIC, which now carries a significant portion of internet traffic~\cite{langley2017quic}. The QUIC protocol's encrypted nature and connection multiplexing capabilities fundamentally alter traditional traffic analysis approaches, requiring new classification methodologies that operate on statistical flow features rather than packet payload inspection. This shift has significant implications for network operators who depend on accurate \kh{TC} for capacity planning, quality of service provisioning, and security monitoring.

Current state-of-the-art QUIC classification approaches rely heavily on centralized machine learning models trained on aggregated network flow data. While recent work has demonstrated that neural networks and ensemble methods can achieve high classification accuracy on QUIC traffic~\cite{tong2018novel,luxemburk2023encrypted}, these approaches assume the availability of comprehensive, centralized datasets spanning multiple network environments. This assumption becomes problematic in real-world deployments where network operators face strict privacy regulations, competitive concerns, and data sovereignty requirements that prohibit sharing raw traffic data across organizational boundaries.

The fundamental tension between the need for comprehensive training data and privacy preservation has led to the exploration of Federated Learning (FL)~\cite{mcmahan2017communication}. Recent work like FedETC~\cite{jin2024fedetc} has already shown that FL is a viable privacy-preserving solution for general encrypted \kh{TC}. However, beyond the initial challenge of privacy, a critical \textit{operational challenge} remains that hinders the practical deployment of FL in real network environments: \textit{temporal traffic volatility}.

Real-world network traffic is not static; it exhibits significant diurnal (daily) fluctuations. This temporal volatility means that clients may have insufficient data during low-traffic periods, leading to noisy model updates that can destabilize the entire synchronous training process. This operational instability, which has been largely overlooked in prior work focused on statistical non-IID data, represents a major barrier to deploying reliable and consistently high-performing FL systems for network management.

This paper is the first to systematically identify and address this stability problem in the context of multi-class QUIC service classification. We propose the novel application of a client-side buffering mechanism—a technique adapted from other FL domains where it served different purposes—to decouple local training from real-time traffic fluctuations and ensure robust model convergence.
Our contributions are:
\begin{itemize}
    \item \kh{We systematically analyze \textit{temporal traffic volatility} as a critical destabilizing factor for synchronous FL in TC}.
    \item We propose and evaluate the application of a \textit{client-side buffering mechanism} as a practical and effective solution to ensure stable model convergence in dynamic network environments.
    \item \kh{We provide a comprehensive evaluation of FL specifically for \textit{multi-class QUIC service classification}, using a realistic 14-client setup and real-network data.}
    \item We demonstrate that our stable federated approach achieves near-optimal performance (95.2\% F1-score), quantifying the small performance trade-off required for a private and operationally robust system.
\end{itemize}

Our experimental evaluation uses the CESNET-QUIC22 dataset~\cite{luxemburk2023cesnet}, partitioned to simulate 14 autonomous clients over a two-week period. Our findings demonstrate that FL, 
when architected with mechanisms like client-side buffering to handle the dynamics of network environments, represents a viable path toward privacy-preserving and \textit{operationally stable}, large-scale QUIC \kh{TC}.

\section{Related Work}
\label{sec:related}

Our research is positioned at the intersection of three key domains: the classification of QUIC traffic, the application of FL to encrypted network data, and the use of stability mechanisms within FL.

\subsection{QUIC Traffic Classification Approaches}

Early research into classifying the then-new QUIC protocol focused on adapting machine learning techniques to its encrypted nature. \citeauthor{tong2018novel}~\cite{tong2018novel} pioneered the use of convolutional neural networks (CNNs), demonstrating that statistical flow features could effectively distinguish QUIC services despite payload encryption. Building on this, \citeauthor{luxemburk2023encrypted}~\cite{luxemburk2023encrypted} conducted a comprehensive comparison of different models using the CESNET-QUIC22 dataset~\cite{luxemburk2023cesnet}, confirming that aggregated flow statistics are highly discriminative. The availability of this large-scale, real-world dataset has been crucial for advancing the field. However, these state-of-the-art approaches all presuppose a centralized data collection model, which presents the significant privacy and logistical challenges that motivate decentralized paradigms like FL.

\subsection{Federated Learning for Encrypted Traffic}

The application of FL to network traffic analysis is an active area of research, primarily driven by privacy needs. Much of the existing work has focused on overcoming challenges related to {statistical} non-IID data. For instance, \citeauthor{guo2021wcl}~\cite{guo2021wcl} and \citeauthor{guo2023feat}~\cite{guo2023feat} proposed sophisticated client selection strategies to improve convergence when data distributions differ across clients. Similarly, \citeauthor{pekar2024incremental}~\cite{pekar2024incremental} explored model performance under extreme non-IID conditions.

Very recently, \citeauthor{jin2024fedetc}~\cite{jin2024fedetc} proposed FedETC, a federated framework for general encrypted \kh{TC}. Their work successfully demonstrates that high-performance, privacy-preserving classification is feasible, achieving accuracy close to that of centralized models. 
However, these existing studies do not address the operational challenge of model stability in the face of real-world traffic fluctuations over time. This temporal volatility is particularly pronounced for QUIC, as many of its dominant applications (e.g., video streaming, social media, real-time communication) are highly interactive and user-driven, leading to pronounced and predictable daily usage cycles compared to background or automated traffic.

\subsection{Buffering Mechanisms in Federated Learning}

The concept of buffering client data or updates is not new to the FL literature, but its application has been targeted at solving system-level and scheduling inefficiencies. For example, \citeauthor{nguyen2021fedbuff}~\cite{nguyen2021fedbuff} introduced FedBuff, a system that uses buffered asynchronous aggregation to improve scalability and tolerate ``straggler'' clients that are slow to respond in large-scale deployments. Other works have used buffers for more efficient client selection or to manage on-device resources. These mechanisms are designed to optimize the FL system's internal operations. To our knowledge, client-side data buffering has not been proposed or evaluated as a mechanism to specifically counteract the destabilizing effect of \textit{temporal data volatility} originating from an external, dynamic process like network traffic.
% 
% In contrast to these system-level optimizations, our application of buffering specifically targets data availability fluctuations caused by external traffic patterns, ensuring consistent gradient quality despite temporal volatility---a challenge unique to network traffic analysis that has not been addressed in prior FL literature.

\subsection{Positioning and Research Gap}

Our work addresses a critical, unaddressed gap at the confluence of these fields. While FL has been applied to encrypted traffic, and buffering has been used for system efficiency, the specific problem of maintaining \textit{stable federated training} for QUIC classification in a dynamic network environment with fluctuating traffic volumes has not been solved. The primary novelty of our paper is therefore not the application of FL to private \kh{TC} in general, but rather the \textit{identification of temporal volatility as a key operational barrier} and \textit{the novel application of a buffering mechanism to solve it}, enabling robust and practical deployment.

\section{Methodology}
\label{sec:methodology}

Our experimental approach evaluates FL for QUIC \kh{TC} through a simulation designed to reflect realistic network operator scenarios. We structure our methodology around three key components: a realistic data distribution that captures both organizational and temporal heterogeneity, a neural network architecture optimized for flow-level features, and a comprehensive FL framework incorporating our proposed stability mechanism.

\subsection{Dataset and Experimental Design}
\label{ssec:dataset}

% Our evaluation uses the CESNET-QUIC22 dataset~\cite{luxemburk2023cesnet}, which contains four weeks of labeled QUIC flows from a backbone network. To create a realistic federated scenario, we collaborated with the dataset's authors to partition the data based on anonymized network prefixes, resulting in 14 distinct clients representing real-world organizational traffic.

\kh{We base our evaluation on the CESNET-QUIC22 dataset~\cite{luxemburk2023cesnet}. It contains labeled QUIC flows collected during four consecutive weeks (Weeks 44--47) from the backbone network infrastructure used by half a million users daily. Since the dataset is anonymized, we contacted the CESNET-QUIC22 authors, who partitioned the data into 14 distinct subsets. Each subset represents traffic from a different IP range associated with a specific, but unknown to us, organization inside the CESNET network.}

%To create a realistic federated scenario, we reached the authors of the dataset, and they partitioned the dataset based on source IP addresses to simulate {14 distinct autonomous clients}, representing different real-world organizations. This partitioning is based on a mapping of network prefixes provided by the dataset's authors, ensuring that each client's data reflects the unique traffic patterns of a single entity.

\kh{Due to the substantial data drift observed in the first two weeks of CESNET-QUIC dataset~\cite{luxemburk2023cesnet}, which could severely affect the experimental results, we limited our analysis to a continuous span of two weeks (Week 46 and Week 47), during which the drift appears to be present but less severe. We further restricted our study to the seven most common and diverse services---Discord, Facebook-Graph, Google-WWW, Instagram, Snapchat, Spotify, and YouTube---to maintain a focused and representative evaluation.}
    
% To mitigate the impact of a known data drift event~\cite{luxemburk2023cesnet} in the dataset's first two weeks, we limit our analysis to a continuous two-week period (Weeks 46 and 47). We further focus on the seven most frequent and diverse services (specifically Discord, Facebook-Graph, Google-WWW, Instagram, Snapchat, Spotify, and YouTube) to ensure a representative and manageable classification task.

%For our experiments, we selected a continuous two-week period\footnote{\kh{We used Week 46 and Week 47}} and focused on the 7 most frequent and diverse services: Discord, Facebook-Graph, Google-WWW, Instagram, Snapchat, Spotify, and YouTube. To analyze the impact of temporal dynamics, we segment the two-week period into 112 consecutive 3-hour intervals.
% Notably, the original dataset spans a period during which Google changed their TLS certificate (November 1, 2022), causing a significant shift in the underlying data distribution. Following the dataset authors' recommendation, we use only the last two weeks of data to ensure consistency.
Each of the 14 clients trains on the data originating from their unique network prefixes within these intervals, naturally exposing the federated system to both {statistical heterogeneity} (different service usage across organizations) and {temporal heterogeneity} (daily traffic fluctuations).

\begin{figure}[t]
\centering
\includegraphics[width=0.48\textwidth, trim=0 0 0 1cm,clip]{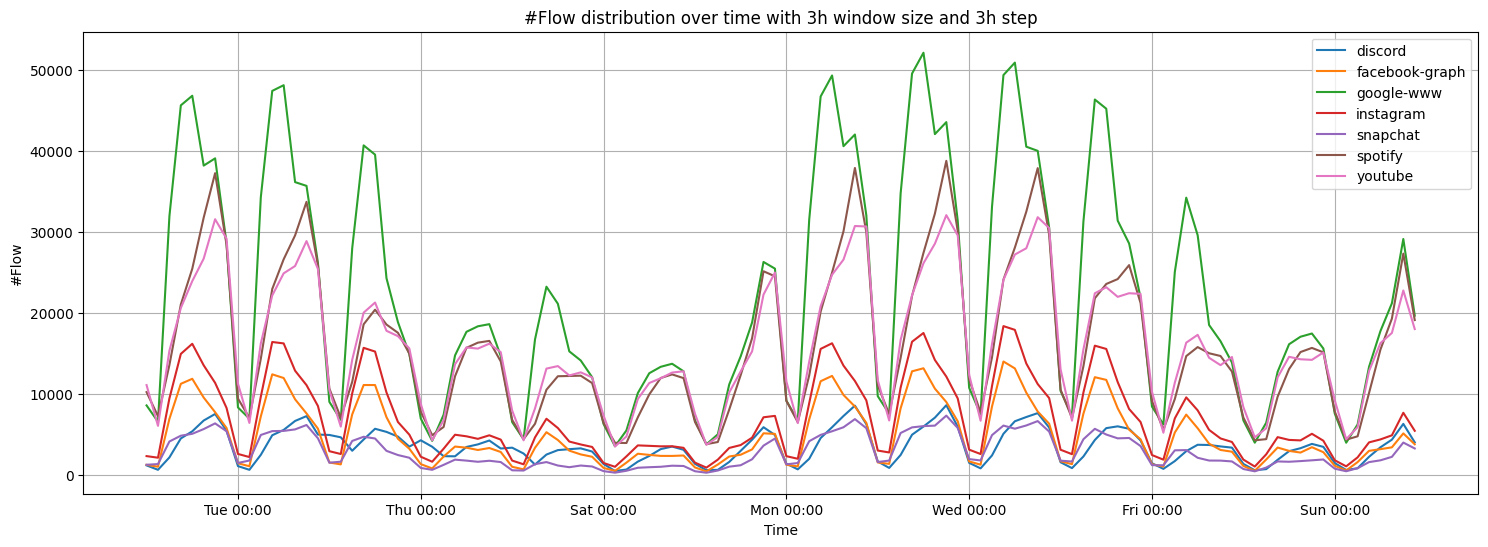}
\caption{\kh{Temporal distribution of services in CESNET-QUIC22 dataset over the selected two-week experimental period.} %, showing pronounced daily usage patterns. 
% The combination of organizational and temporal partitioning creates a uniquely realistic non-IID benchmark for FL.
}
\label{fig:traffic_distribution}
\end{figure}

The distribution of flows across our 3-hour intervals, illustrated in \Cref{fig:traffic_distribution}, reveals pronounced daily cycles common to real networks. This setup provides a powerful benchmark for evaluating the robustness of FL algorithms under practical, challenging conditions.

\subsection{Feature Engineering}
\label{ssec:features}

The raw CESNET-QUIC22 dataset provides basic flow statistics and packet-level information (PPI) for the first 30 packets. To create a more discriminative feature set for classification, we perform two key feature engineering steps:

\begin{itemize}
    \item \textit{Per-Direction Packet Features}: We disaggregate the initial packet information (size and inter-arrival time) into separate features for each communication direction (client-to-server and server-to-client). This captures the asymmetric nature of client-server interactions.
    \item \textit{Per-Direction Flow Statistics}: From these directional packet features, we compute higher-level statistics, such as the mean, standard deviation, min, and max of packet sizes and inter-arrival times for each direction.
\end{itemize}

This process results in a comprehensive feature vector of $N=283$ features for each flow, combining the original dataset's statistics with our engineered directional features. All features are scaled to a [0, 1] range using min-max scaling before being fed into the neural network.

\subsection{Neural Network Architecture}
\label{ssec:architecture}

Our classification model employs a fully connected neural network (FCN) designed for tabular flow statistics. The network uses an expanding structure to learn progressively complex feature interactions:

\begin{itemize}
    \item \textit{Input Layer}: $N$ neurons (number of input features)
    \item \textit{Hidden Layer 1}: $2N$ neurons + BatchNorm + LeakyReLU(0.1) + 15\% Dropout
    \item \textit{Hidden Layer 2}: $3N$ neurons + BatchNorm + LeakyReLU(0.1) + 15\% Dropout
    \item \textit{Hidden Layer 3}: $3N$ neurons + BatchNorm + LeakyReLU(0.1) + 15\% Dropout
    \item \textit{Hidden Layer 4}: $4N$ neurons + BatchNorm + LeakyReLU(0.1) + 15\% Dropout
    \item \textit{Output Layer}: 7 neurons (one for each service class)
\end{itemize}

Hidden layers incorporate batch normalization for training stability, LeakyReLU activation to prevent the dying ReLU problem, and 15\% dropout for regularization. The model is trained using cross-entropy loss and the Adam optimizer with a learning rate of 0.001 and batch size of 64 for federated clients (compared to 0.01 and 1024 for centralized training) to ensure more robust gradient computation with limited local data. Each local training round consists of 10 epochs.

\subsection{Federated Learning Configuration}
\label{ssec:fl_config}

Our FL simulation utilizes the Flower framework~\cite{beutel2020flower} with PyTorch. The protocol follows standard FL communication rounds where the server broadcasts the global model, clients train locally, and return updated parameters. We systematically evaluate five aggregation algorithms to understand their behavior in this context:
\begin{itemize}
    \item \textit{FedAvg}~\cite{mcmahan2017communication}: Baseline weighted parameter averaging.
    \item \textit{FedProx}~\cite{li2020fedprox}: Adds a proximal term to handle client drift.
    \item \textit{FedAdam}, \textit{FedAdagrad}, \textit{FedYogi}~\cite{reddi2020adaptive}: Implement server-side adaptive optimization.
\end{itemize}

% For FedAvg, client updates are weighted by the number of local training samples. The FedProx proximal term weight is set to 0.1, while adaptive optimizers use $\beta_1=0.9$, $\beta_2=0.99$, with learning rates of 0.1 (0.01 for FedYogi).

\subsection{Addressing Temporal Volatility with Client-Side Buffering}
\label{ssec:buffer}

A key challenge we identified in applying synchronous FL to network traffic is the model instability caused by temporal data volatility. To address this, we propose a novel application of a \textit{client-side FIFO (First-In, First-Out) data buffer}. 
% While the principles of buffering have been explored in the FL literature for different purposes, such as handling system stragglers~\cite{nguyen2021fedbuff}, we are the first to apply this technique to counteract the effects of fluctuating data availability from an external process like network traffic.
% 
In our framework, instead of training on all data collected in a 3-hour interval, each client accumulates incoming flow records into a fixed-size buffer. A local training round is only initiated once the buffer contains a sufficient number of samples (6400 records in our experiments). If new data causes the buffer to exceed its capacity, the oldest records are discarded. 

This mechanism serves to decouple local model training from real-time traffic volume. By ensuring that each local update is derived from a consistent and statistically significant amount of data, the buffer smooths out the impact of temporal fluctuations, leading to more stable and robust global model convergence. This directly addresses the temporal volatility challenge identified as a key barrier to practical deployment.

\subsection{Client-Side Data Handling}
\label{ssec:data_handling}

To evaluate the impact of temporal volatility, we define two distinct client-side data handling strategies:
\begin{itemize}
    \item \textit{Standard FL (unbuffered):} In each 3-hour round, clients use all data that arrived in that interval. This data is split into training (70\%), validation (10\%), and test (20\%) sets. This results in highly variable dataset sizes per round, directly reflecting traffic volatility.
    \item \textit{Buffered FL (our proposal):} \kh{Clients maintain three fixed-size FIFO buffers for training (6400 flows), validation (914), and testing (1828), reflecting the selected 70/10/20 ratio. In each FL round, newly arriving flows are partitioned according to this ratio and added to the respective FIFO buffers. This approach ensures stable and consistent dataset sizes for training and evaluation across rounds.}
\end{itemize}
In both scenarios, the validation set is used for local early stopping, and the test set is used to report the performance for each round.

\subsection{Evaluation Framework}
\label{ssec:evaluation}

% Our framework assesses both classification performance and FL dynamics. The primary metric is the macro F1 score, which provides a balanced evaluation for multi-class scenarios. 

% We first establish a centralized baseline by training the neural network on the complete, aggregated dataset. This provides a performance upper bound to quantify the privacy-performance trade-off of our federated approach.

% Feature importance is assessed using SHAP (SHapley Additive exPlanations)~\cite{lundberg2017unified} to validate that the model learns meaningful patterns on the centralized model. 
% This analysis aims to identify the most fundamentally predictive flow features under ideal conditions, providing a ground truth for interpreting the behavior of the federated system.

% Beyond final performance, we analyze the convergence behavior across the 112 federated rounds. We specifically compare the stability of the system \textit{with and without our proposed buffering mechanism} to quantify its impact on mitigating temporal volatility. 

Our framework assesses both classification performance and FL dynamics. We first establish a \textit{centralized baseline} using the complete, aggregated dataset to define an upper performance bound. 

We use SHAP (SHapley Additive exPlanations)~\cite{lundberg2017unified} on the centralized model to identify the most predictive flow features under ideal conditions, providing a ground truth for interpreting the model's behavior.

Finally, for the federated scenarios, we directly compare the convergence stability of the \textit{standard (unbuffered) FL} against our \textit{proposed buffered FL} to quantify the impact of our mechanism. 
% The primary performance metric is the weighted macro F1 score.
The primary performance metric reported in this paper is the macro F1 score; however, additional metrics are provided in the digital artifact~\cite{GitHub}.

\section{Experimental Results}
\label{sec:results}

% Our evaluation is structured to first establish a performance baseline, then demonstrate the stability challenge of standard FL, and finally, show how our proposed mechanism solves this challenge to achieve near-optimal performance.

\subsection{Centralized Performance and Feature Insights}
\label{ssec:results_centralized}

\kh{We first establish an upper-bound performance by training our FCN on the complete, aggregated two-week dataset. This \textit{centralized baseline} achieves a {97.5\% F1 score}. The classifier showed strong per-class performance, where all classes were recognized with more than 95\% precision, confirming the general separability of the QUIC services with our engineered features. Discord and Spotify are classified with over 99\% precision, indicating highly distinctive traffic patterns. Social media services like Instagram (96.74\%) and Facebook-Graph (95.26\%) show more confusion, likely due to overlapping backend APIs and content delivery networks.}

% \begin{figure}[t]
% \centering
% \includegraphics[width=0.48\textwidth]{figures/confusion_matrix.png}
% \caption{Confusion matrix for the centralized model. 
% % Discord achieves the highest precision (99.93\%), while Facebook-Graph shows the most confusion with other services. The overall diagonal dominance indicates strong separability of QUIC services using our engineered flow-level features.
% }
% \label{fig:confusion_matrix}
% \end{figure}

\begin{figure}[t]
\centering
\includegraphics[width=0.4\textwidth, trim=0 0 0 1.2cm, clip]{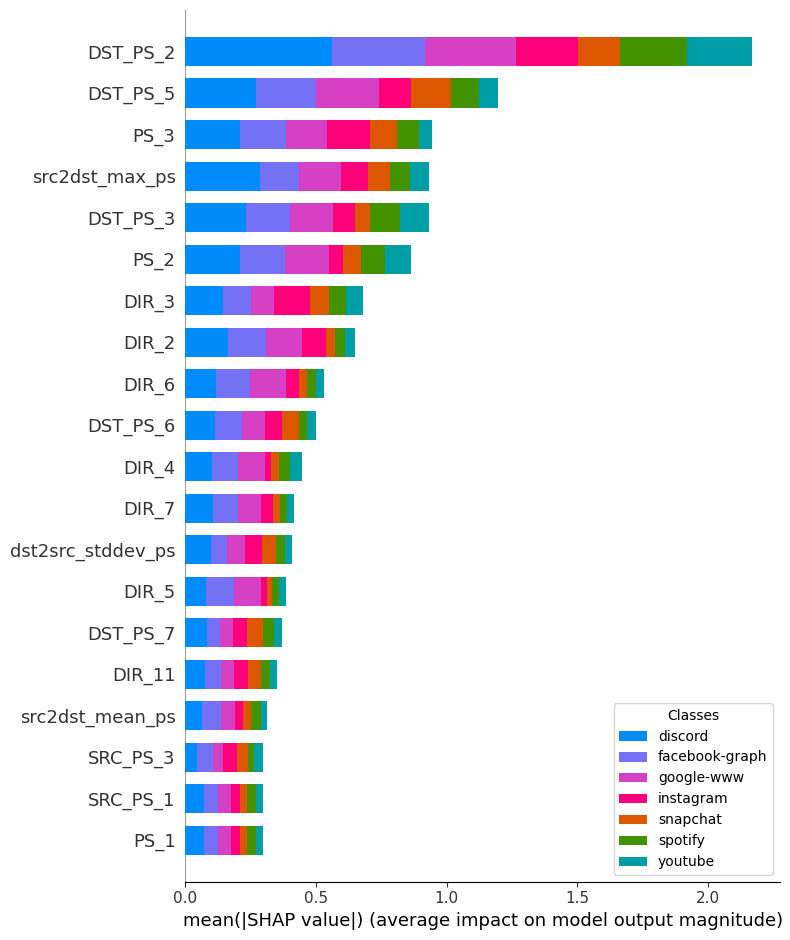}
\caption{SHAP analysis on the centralized model identifies the most influential flow features for QUIC classification. \kh{\texttt{PS\_n} denotes the size of the \textit{n}-th packet in the bidirectional flow sequence, while \texttt{DST\_PS\_n} and \texttt{SRC\_PS\_n} specifically indicate server-sent and client-sent packets, respectively}
% Packet size statistics from the server (e.g., `DST\_PS\_2`) are the most influential, confirming the importance of server-side traffic shaping in identifying services.
}
\label{fig:shap_analysis}
\end{figure}

To understand which features are fundamentally most important, we performed a SHAP analysis on this centralized model. The results in \Cref{fig:shap_analysis} reveal that packet size statistics from the server dominate the rankings. Specifically, `DST\_PS\_2` is the most influential feature. This aligns with domain knowledge, as this packet often contains the server's TLS certificate, whose size is a highly discriminative service signature. 
% Our feature engineering revealed that directional packet features alone (SUBPSTATS) achieve 96.6\% F1 score, nearly matching the baseline while requiring significantly less computational complexity than full flow statistics. 
This analysis provides a ground truth for what an ideal model should learn.

\subsection{Instability of Standard FL}
\label{ssec:results_without_buffer}

\begin{figure*}[t]
    \centering
    % Subfigure for "Without Buffering"
    \begin{subfigure}[b]{0.48\textwidth}
        \centering
        \includegraphics[width=\textwidth]{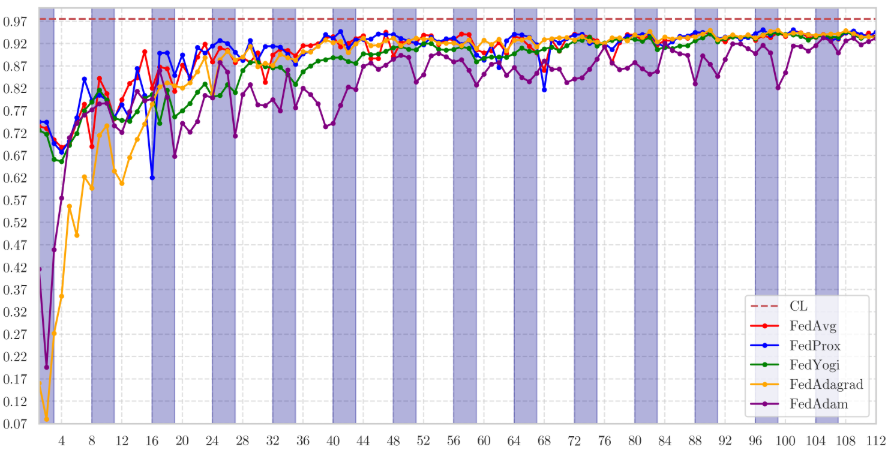}
        \caption{Without buffering (Standard FL)}
        \label{fig:convergence_without_buffer}
    \end{subfigure}
    \hfill
    % Subfigure for "With Buffering"
    \begin{subfigure}[b]{0.48\textwidth}
        \centering
        \includegraphics[width=\textwidth]{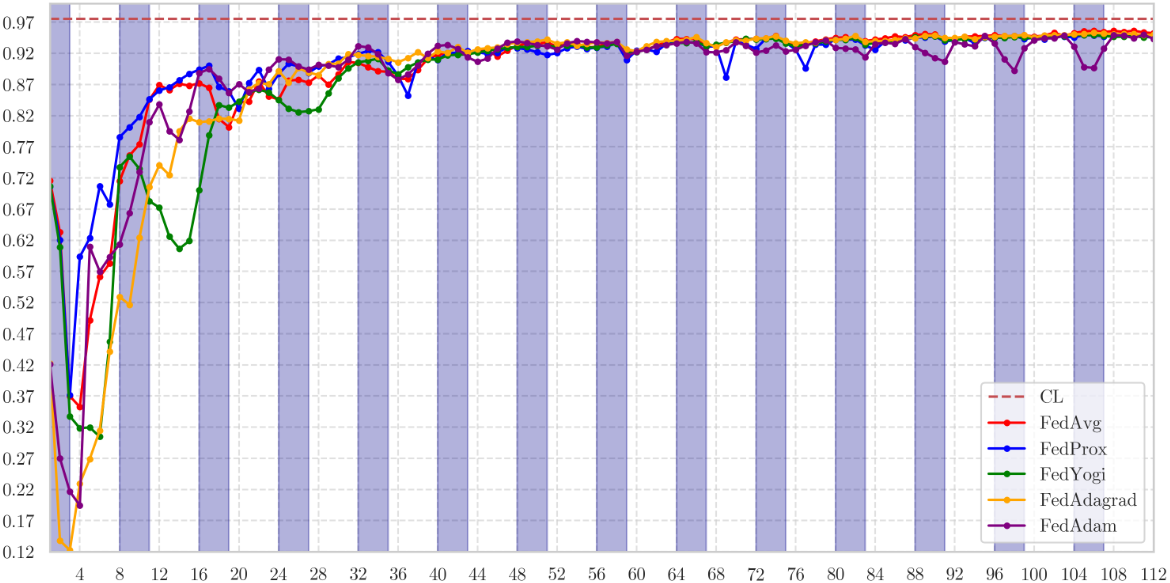}
        \caption{With our proposed buffering mechanism}
        \label{fig:convergence_with_buffer}
    \end{subfigure}
    
    \caption{Comparison of FL convergence behavior over 112 rounds (two weeks). 
    % Shaded regions indicate low-traffic nighttime periods. 
    % (a) Without our stability mechanism, performance is highly volatile and synchronized with daily traffic cycles. (b) With client-side buffering, performance stabilizes, demonstrating the effectiveness of our approach in mitigating temporal volatility.
    }
    \label{fig:convergence_comparison}
\end{figure*}

Next, we evaluate a standard synchronous FL setup where clients train only on the data available in each 3-hour round. \Cref{fig:convergence_without_buffer} shows the result: the system is plagued by instability. The performance of all five aggregation algorithms plummets during low-traffic nighttime periods (shaded regions), with some clients (particularly clients 10 and 14 in our experiments) failing to converge entirely due to insufficient data for reliable gradient estimation. The dashed red line represents the centralized baseline F1 score. It is clear that without a stability mechanism, standard FL consistently and significantly underperforms.

\subsection{Stable and High-Performance Buffered FL}
\label{ssec:results_with_buffer}

Finally, we evaluate our proposed FL framework with the client-side buffering mechanism enabled. \Cref{fig:convergence_with_buffer} demonstrates the dramatic improvement. The introduction of the buffer almost completely eliminates the volatility, and all algorithms achieve stable convergence.

% Consistent with the unbuffered experiment, simple parameter averaging (FedAvg) outperforms more sophisticated algorithms in this domain. 

FedAvg achieves both the highest average performance and the best stability (lowest standard deviation) through simple weighted averaging based on client data sizes, outperforming more complex adaptive optimizers that proved sensitive to gradient noise. Notably, FedAvg with buffering, achieves a stable F1 score of 95.2\%. This represents a performance trade-off of only 2.3 percentage points compared to the non-private centralized model. 
This suggests that in the presence of non-stationary network traffic, the robustness of a simple average is more beneficial than the complex, and often unstable, updates of adaptive methods like FedAdam. 
% The pronounced performance degradation of FedAdam in the final rounds (rounds 100-112) exemplifies the documented instability of adaptive optimizers in federated settings, where the momentum-based second moment estimation becomes increasingly unreliable as heterogeneous gradient patterns from different clients cause the adaptive learning rates to oscillate rather than converge smoothly~\cite{shin2024not}.
This instability of adaptive optimizers in heterogeneous federated settings is a documented phenomenon, as their momentum-based updates can fail to converge smoothly~\cite{shin2024not}.

\section{Conclusion}
\label{sec:conclusion}

In this paper, we addressed the critical challenge of classifying encrypted QUIC traffic in a scalable and privacy-preserving manner. We demonstrated that FL, when properly adapted to the unique dynamics of network environments, is a highly effective solution. Our proposed approach, combining the simplicity of the FedAvg algorithm with a novel client-side buffering mechanism, achieves an F1 score of 95.2\%. This represents a remarkably small performance degradation of only 2.3 percentage points compared to an ideal, non-private centralized model.

The experimental results carry practical implications for network operators. The 2.3\% performance cost for privacy is a highly favorable trade-off, making collaborative, cross-organizational traffic analysis feasible without compromising data sovereignty. Furthermore, our findings strongly advocate for simplicity in system design; the combination of a basic FedAvg aggregator with a client-side stability mechanism like our buffer provides a robust, effective, and computationally efficient solution for real-world deployment on network monitoring platforms.

Our work opens several avenues for future research. We suggest exploring \textit{1)} asynchronous FL to offer greater client flexibility; \textit{2)} sequential models (e.g., LSTMs) to improve accuracy on long-lived flows; and \textit{3)} adaptive buffering strategies to dynamically optimize the trade-off between data freshness and model stability.

% Our key contribution is twofold. First, we identified that the natural, daily volatility of network traffic is a primary source of instability for standard FL algorithms, particularly for adaptive methods. Second, we showed that this instability can be effectively mitigated through a simple client-side FIFO buffer that decouples local training from real-time data availability. This finding underscores a crucial principle for practical FL deployment in networking: system stability mechanisms are as important as the choice of aggregation algorithm. Our work provides a concrete blueprint for building robust, privacy-preserving traffic analysis systems that can be deployed across multiple, competing network operators without requiring them to share sensitive data.

\section*{Reproducibility}
To ensure the reproducibility of our results and to facilitate future research, all source code, experimental configurations, and data processing scripts are publicly available at our GitHub repository~\cite{GitHub}. Beyond the results presented in this paper, the artifact provides extended analyses, metrics, and complementary visualizations, offering deeper insights into per-client performance and the temporal dynamics of the system.

\section*{Acknowledgment}
Supported by the János Bolyai Research Scholarship of the Hungarian Academy of Sciences.
% This work has been part of Celtic-Next project RAI-6Green: Robust and AI Native 6G for Green Networks with project-id: C2023/1-9 funded by 2024-1.2.6-EUREKA-2024-00009.
This work was also supported by projects TKP2021-NVA-02 (financed under the TKP2021-NVA scheme) and 2024-1.2.6-EUREKA-2024-00009 (financed under the 2024-1.2.6-EUREKA scheme), implemented with support provided by the Ministry of Culture and Innovation of Hungary from the National Research, Development and Innovation Fund.

\printbibliography

\end{document}